\documentstyle[prb,aps,multicol,epsf]{revtex}
\twocolumn
\begin{document}
\title{Conductance of a Quantum Point Contact in the presence of
a Scanning Probe Microscope Tip}
\author{Guang-Ping He$^{1}$, Shi-Liang Zhu$^{2,3}$, 
and Z. D. Wang$^{2,4,5}$}
\address{$^{1}$Advanced Research Center, Zhongshan University, Guangzhou 510275, China\\
$^{2}$Department of Physics, University of Hong Kong, Pokfulam Road,
Hong Kong, China\\
$^{3}$Department of Physics, South China Normal University,
Guangzhou 510631, China\\
$^{4}$ Department of Material Science and Engineering, University of Science and Technology of China, Hefei, China\\
$^{5}$ Texas Center for Superconductivity, University 
of Houston, Texas 77204, USA
}
\address{\mbox{}}
\date{Phys. Rev. B 65, 205321 (2002)}
\address{\parbox{14cm}{\rm \mbox{}\mbox{}
Using  the recursive Green's function technique, we study
the coherent electron conductance of a quantum point contact 
in the presence of a scanning probe microscope tip.
Images of the coherent fringe
inside a quantum point contact for different widths
are obtained.
It is found that
the conductance of a specific channel is
reduced while other channels are not affected as long as
the tip is located at the 
positions correspending to that channel.
Moreover, the coherent fringe
is smoothed out
by increasing the temperature or the voltage across
the device.
Our results are consistent with the experiments reported by
Topinka {\it et al.}(Science 289, 2323 (2000)).
}}
\address{\mbox{}}
\address{\parbox{14cm}{\rm PACS numbers: 73.23.-b, 73.43.Cd, 73.20.At}}

\maketitle

\newpage

Quantum point contacts (QPCs) formed in two-dimensional electron gases
(2DEGs) have attracted significant attention
for the past two decades.\cite{Wees,Beenakker}
Since the discovery of the conductance quantization
in these structures,\cite{Wees}
QPCs have been widely used in a variety of investigations, including transport
through quantum dots, the quantum Hall effect, magnetic focusing, and the
Aharonov-Bohm effect. \cite{Beenakker}
Also, with the rapid
development on scanning probe microscope (SPM) techniques, it is
possible to image current directly to study many remarkable phenomena,
including quantum corrals,\cite{Crommie}
electron flow through nanostructures,\cite{Eriksson_96}
charge distribution and photoactivity of dopant atoms,\cite{Yoo}
and spectra of metallic nanoclusters.\cite{Gurevich}
Since the QPC plays such an
important role in mesoscopic devices,
it is an ideal system
to be studied by the SPM techniques.

Very recently, Topinka {\sl et al}. 
have directly imaged the electron flow from
the QPC by scanning a negatively charged
SPM tip above the surface of
the device and measuring the position-dependent
conductance simultaneously.\cite{Topinka_S,Topinka_N}
In the experiment, as the width of the QPC
increases, the conductance increases in quantized steps
of $2e^2/h$,
as was also reported in other experiments.\cite{Wees}
Besides, several new and interesting features were observed.
Widening of the angular structure of electron flow
has been clearly seen
as the QPC channel becomes wider.
Another feature of their images is the appearance of fringes
spaced by half the Fermi wavelength transverse to the
electron flow, which clearly shows the character
of coherent quantum interference.
The above fringes are thermally smeared out by increasing
the effective electron temperature.
Moreover, when the tip is placed such that it interrupts the
flow from particular modes of the QPC,
a reduction can be observed in the
conductance of those channels, while
other channels are not affected.\cite{Topinka_S}
On the other hand,  it is also found that,
in contrary to intuitions, the
electron flow from the point contact forms narrow
branching strands instead
of smoothly spreading fans.\cite{Topinka_N}

Using the recursive Green's function technique,
in this paper, we  
study theoretically a mesoscopic structure similar to the
experimental one.\cite{Topinka_S,Topinka_N}
Our numerical results are able to explain the
experimental finding in Refs.7,8.
Therefore, the recursive Green's function technique
seems to be a suitable method to study the
coherent electron conductance in such a mesoscopic system.

\begin{figure}[tbp]
\vspace{-0.5cm}
\label{Fig1}
\epsfxsize=7.5cm
\epsfbox{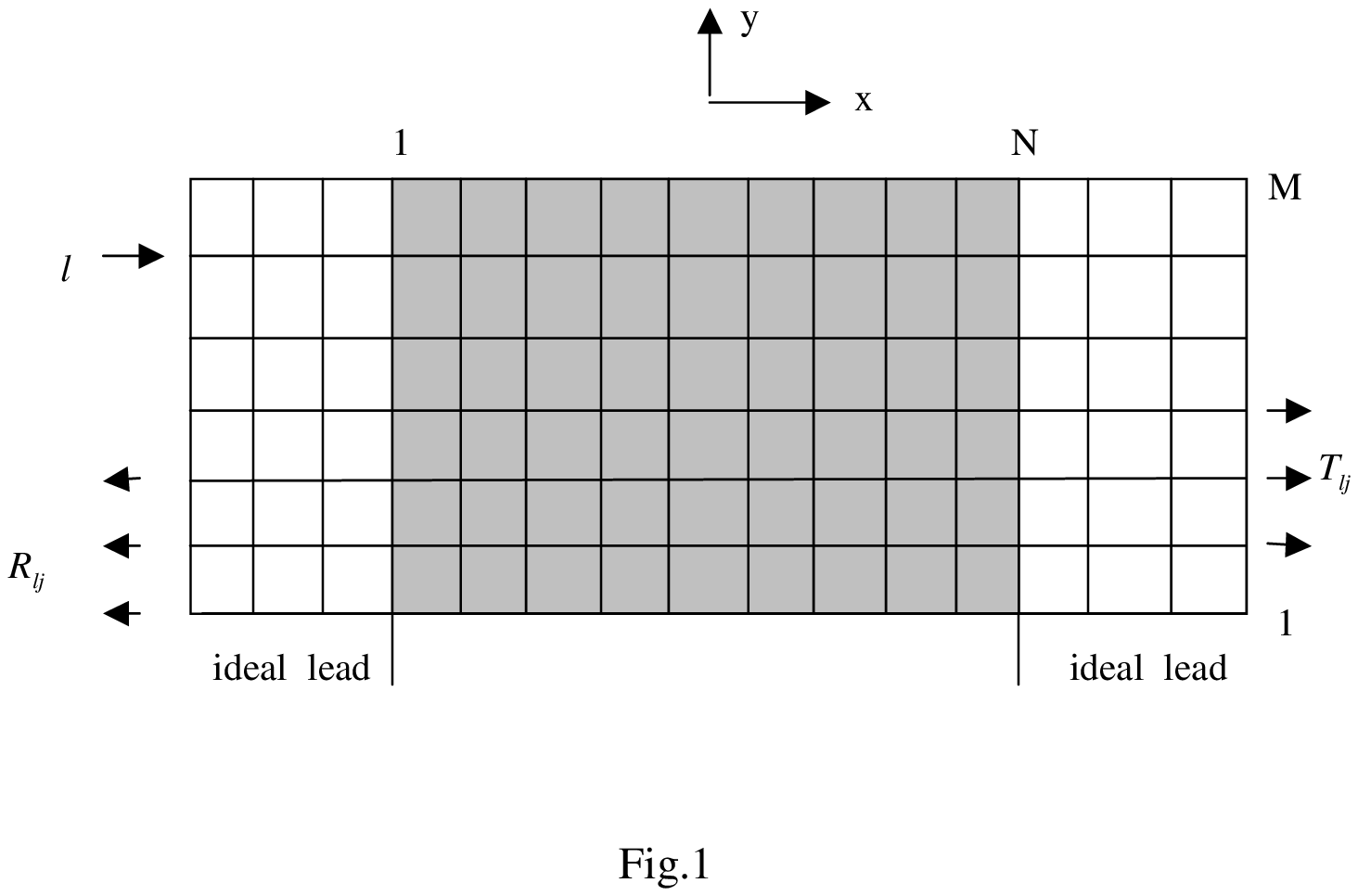}
\vspace{-4.0cm}
\caption{Schematic diagram of the system.}
\end{figure}

We consider a negatively charged SPM tip capacitively coupled to
the 2DEG where
the QPC is formed. 
The effect of SPM tip is that,
the conductance will decrease if the tip is positioned over regions
with high electron flow from the QPC, while it will be less modified if the
tip is at low electron flow regions.
The electron flow can be imaged by
scanning the tip over the two-dimensional
surface.
Two contributions to the potential in the 2DEG region
should be taken into account. One is from
the negatively charged gates that define the QPC, which we model with
the B\"{u}ttiker saddle-point potential
$U(x,y)=V_{g}-m^{\ast}\omega_{x}^{2} x^{2}/2+
m^{\ast}\omega _{y}^{2} y^{2}/2$ 
with $V_g$ as the gate voltage,
$m^*$ as the electron effective mass and
$\omega_{x,y}$ as the strength of the lateral
confinement.
This potential is a practical candidate to 
reproduce the quantized conductance
of a QPC.\cite{Buttiker}
The other is a Lorentzian type potential
induced by the SPM tip,\cite{Eriksson} 
 which we approximate 
as a delta function  in the present work : 
$V_{SPM}=V_{t}\delta (\overrightarrow{r}-\overrightarrow{r_{t}})$,
where $\overrightarrow{r_{t}}$ denotes the position of the tip, and $V_{t}$
is a negative constant.
It is worth  noting that
the numerical results obtained with the simple delta potential
are consistent the experimental results.\cite{Topinka_S,Topinka_N}

We use a 2D square lattice model to describe the QPC system.
The sites of the lattice are denoted as $(na,ma)$
with $a$ to be set equal to unit, $%
n=1,2,...,N $\ and $m=1,2,...,M$.
The lattice may be divided into three regions,
as shown in Fig.1. The shadowed central zone is a mesoscopic structure with
a B\"{u}ttiker saddle-point potential and scanned by a SPM tip. 
Both sides of the structure are assumed to be
connected with semi-infinite ideal leads to simplify scattering boundary
conditions. The one-electron tight-binding Hamiltonian of the system
takes the form \cite{Zhu}
\begin{eqnarray}
\nonumber
H &=&\sum\limits_{n,m}\varepsilon _{n,m}\left| n,m\right\rangle \left\langle
n,m\right|\\
&-&\sum\limits_{n,m}(t| n,m\rangle
\langle n-1,m | +t| n,m\rangle \langle n,m-1| +H.C.),
\label{eq3}
\end{eqnarray}
where
$t=\hbar^{2}/2m^{\ast }a^{2}$
and $|n,m\rangle$ is a orthonormal set in the
lattice sites $(n,m)$. The on-site energy
in the central zone is given by
$\varepsilon _{n,m}=U_{n,m}+V_{SPM}+4t$.
Then the
Hamiltonian for this region reads 
\begin{equation}
\label{Hamiltonian_C}
H_{c}=\sum\limits_{n=1}^{N} |n) H_{n} (n|
+\sum\limits_{n=1}^{N-1}( |n) H_{n,n+1} (n+1| +H.C.),  
\end{equation}
where $ |n) $ denotes the set of $M$ ket vectors belonging
to the $n$-th cell, and 
$$
H_{n}\equiv H_{n,n}=\left[ 
\begin{array}{ccccc}
\varepsilon _{n,1} & -t & 0 & \cdots & 0 \\ 
-t & \varepsilon _{n,2} & -t & \cdots & 0 \\ 
0 & -t & \varepsilon _{n,3} & \cdots & 0 \\ 
\vdots & \vdots & \vdots & \ddots & \vdots \\ 
0 & 0 & 0 & \cdots & \varepsilon _{n,M}
\end{array}
\right] ,  
$$
$$
(H_{n,n-1})_{pp^{\prime }}=(H_{n,n+1})_{pp^{\prime }}=-t\delta _{pp^{\prime }}
\ \ (p,p^{\prime }=1,...,M). 
$$
The Hamiltonians for the two ideal leads have the same
form as Eq.(\ref{Hamiltonian_C}),
but with different summing regions ($-\infty <n<1$ at the left lead and $%
N<n<\infty $ at the right lead) and $\varepsilon _{n,m}=4t$.

Using the recursive Green's function technique, 
the transmission amplitude for the incident channel
$l$ and outgoing channel $j$
is found to be\cite{Zhu,Ando}
\begin{equation}
\label{coefficient}
t_{lj}
=\sqrt{v_j/v_l}\phi^+_{k_j}
G_{N+1,0}\Theta(E) \phi_{k_l}
e^{-ik_j L_x},
\end{equation}
where $ v_j={\partial E}/\hbar{\partial k_j}$
with $E=4t-2t(cosk_j+cos\pi j/(M+1))$,
$ \Theta(E)=-2it\sum_{l=1}^{M} Q_l sink_l$ with
$$
(Q_l)_{pp'}=\frac{2}{M+1}sin\frac{l\pi p}{M+1}sin\frac{l\pi p'}{M+1},
\ (p,p'=1,\cdots,M).
$$
The set of vectors $\phi^+_{k_j}$
are the duals of the set $\phi_{k_l}$, defined by
$\phi^+_{k_j} \phi_{k_l}
=\delta_{jl}$ where
$$
\phi _{k_{j}}=\sqrt{\frac{2}{M+1}}(\sin \frac{\pi j}{M+1},...,\sin \frac{\pi
jm}{M+1},...,\sin \frac{\pi jM}{M+1})^{T_r}
$$
with $T_r$ as the transposition of matrix.
$G_{N+1,0}$ is the retarded Green's function for
the scattering region between two ideal
leads, which can be obtained by a set of
recursion formulas in a matrix form,~\cite{Ando,Lee}
\begin{eqnarray}
\label{Gfunction}
G_{n'+1,0} &=& -tg^{n'+1}G_{n',0},\ \ (0\leq n' \leq N) \\
g^{n'+1} &=& (E-\widetilde{H}_{n'+1}-t^2g^{n'})^{-1},  
\end{eqnarray}
by iteration starting form
$g^0=G_{0,0}=(E-\widetilde{H}_0)^{-1}$, where
$\widetilde{H}_l=H_l\ (1\leq l \leq N)$,
and $\widetilde{H}_{\alpha}=H_{\alpha}-
t\sum_{j=1}^{M} e^{ik_j} Q_j$ $(\alpha=0,N+1)$.

At finite temperature $T$, the conductance through a 2D mesoscopic
structure is given by the Landauer-B\"{u}ttiker formula\cite{Landauer} 
\begin{equation}
\label{eq21}
G(T)=\frac{2e^{2}}{h}\sum\limits_{l,j}
\int\nolimits_{0}^{\infty }T_{lj}\frac{\partial
f(E,T)}{\partial E}dE,  
\end{equation}
where $T_{lj}=| t_{lj}| ^{2}$
is the transmission coefficient,
and $f(E,T)=[1+\exp (E-E_{F})/k_{B}T]^{-1}$ is the
Fermi-Dirac distribution with $E_{F}$ as the Fermi energy and $k_{B}$ as the
Boltzmann constant.
Obviously, at zero temperature Eq.(\ref{eq21}) reduces to 
$G=2e^{2}/h\sum_{ \{ l,j \} }T_{lj}$.

The above formulas are valid under the assumption that the voltage across
the device is low. High voltage will raise the energies of electrons flowing
from the QPC. Thus, as shown by van Wees {\it et al.}, \cite{Wees_91} the
conductance at a finite voltage $V$  has a similar behavior to that
with a finite temperature effect, and is given by 
\begin{equation}
G(V)=\frac{2e}{h}\frac{1}{V}\sum\limits_{l,j}
\int\nolimits_{E_{F}}^{E_{F}+eV}T_{lj}dE.
\end{equation}

In the following we calculate numerically the conductance of a QPC system
with the size $N=15$ and $M=11$. We set $t$
and the lattice constant $a$ as the units of energy and length,
respectively. In order to obtain well-pronounced quantized plateaus with the
B\"{u}ttiker saddle-point potential,
we choose $\omega _{y}=2\omega _{x}$,
$E_{F}=2t$, and $V_{t}=-1.5t$.

\begin{figure}
\label{Fig2}
\epsfxsize=7.5cm
\epsfbox{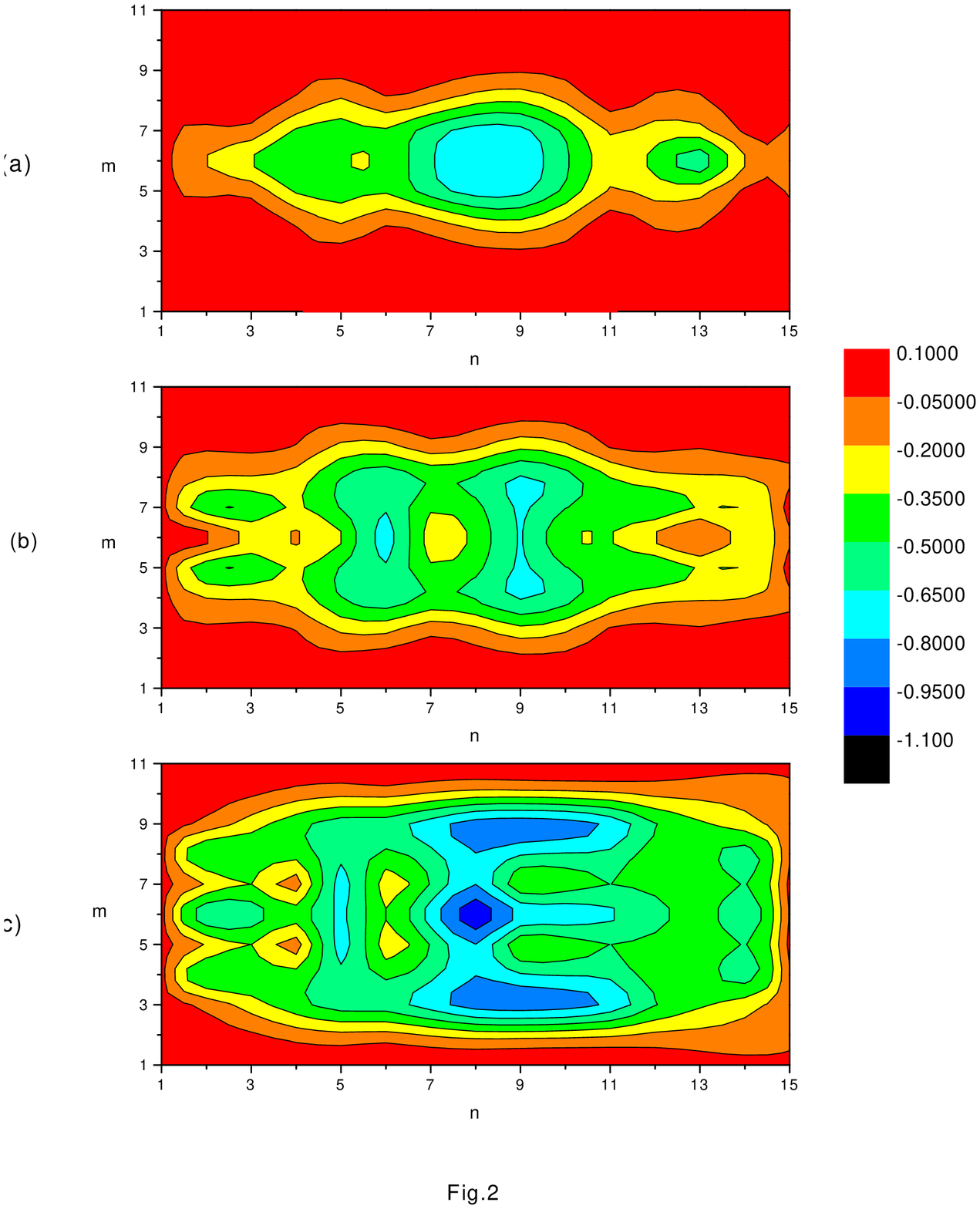}
\caption{
Images of electron flow
across a QPC at zero temperature and low voltage
for three different widths corresponding to
(a) the first conductance plateau, $V_{g}=0.7E_{F}$;
(b) the second plateau, $V_{g}=0.4E_{F}$;
(c) the third plateau, $V_{g}=0.2E_{F}$.
The color scale shows the
change $\Delta G=G-G_{0}$ in QPC conductance
as the tip is scanned above the device.
}
\end{figure}

At zero temperature and low voltage limit, images of electron flow from the
2D surface of the QPC are plotted in Fig.2. The gate voltage $V_{g}$ is
chosen to ensure Fig.2(a) - (c) correspond to the
first, the second and the third
conductance plateau of the QPC, respectively.
It is seen clearly that
a new electron flow pattern appears
when a higher mode is opened (then the conductance of the QPC rises to
another plateau) by decreasing the gate voltage.
This feature agrees with the result reported in Ref.7:
electronic wavefunctions inside the QPC
have $N$ maxima when QPC has $N$ conductance modes.
Note that
the images do not turn into
smoothly spreading fans
as the electrons flow
across the QPC from the left to the right.
Instead, they form branching strands which are relatively
narrower. In fact, even the width of the sample decreases
to $M=6$ in the calculation,
the images change little (not shown here).
It is not surprising
that strands are formed in our model
since the electrons are laterally confined.
On the other hand, a similar but also striking phenomenon is
reported in Ref.8, 
where the strands were found in regions far away
from the point contact .
So we can see that the electron path has a rare probability
to pass through the
edge regions of the sample, and thus
the impurity density and the boundary scattering
in these regions are less relevant to the
conducting property of the QPC. Moreover, the fringe structures
due to the alternating constructive and destructive
interference of electrons are evident in Fig.2, though
the distribution is not so homogeneous
when compared with the experiments.\cite{Topinka_S}
This inhomogeneity
induced from the varieties of electron wavelength
inside the QPC is the main difference between
the structure of fringe inside and outside QPC.
It is worth pointing out that
the fringes directly demonstrate
the coherent character of electron flow in QPC.

\begin{figure}
\label{Fig3}
\epsfxsize=7.5cm
\epsfbox{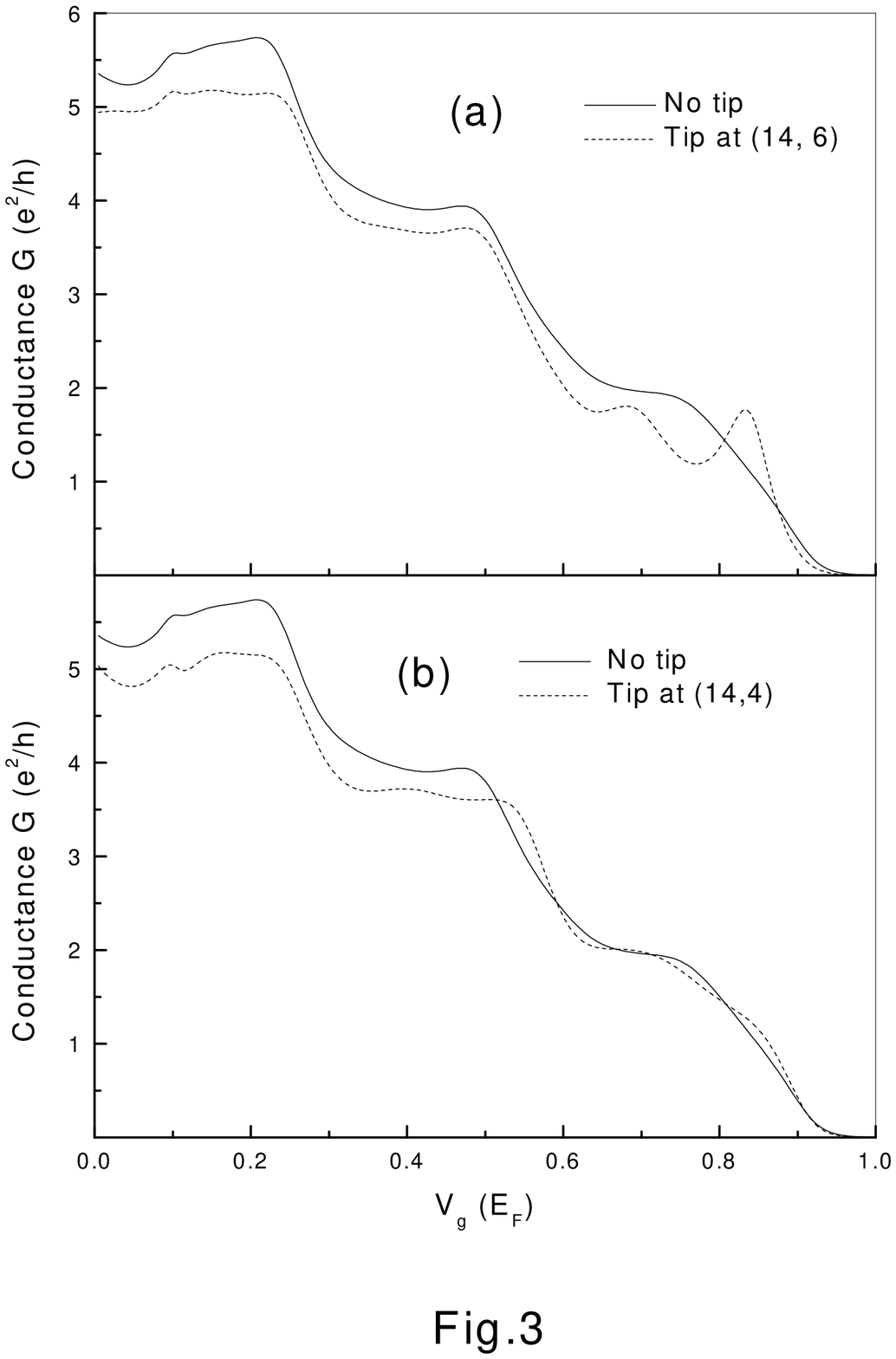}
\caption{
The conductance plateaus of the QPC
for the selective effect of the SPM tip. The solid line
represents the curve without SPM tip,
and the dash line is the
curve with the SPM placed over certain position such that it blocks the
electron flow from (a) the first mode, or (b) the second mode.
}
\end{figure}

Meanwhile, we also note that the
angular structures in these electron flow patterns are
quite different from each other, especially for the
first and second modes. For the first mode, the pattern contains one
branch only, while there are two branches for the second mode.
So there are
some regions in the 2D surface where the
distribution of electrons at one mode make no
contribution to the other mode. As reported in
experiments,\cite{Topinka_S} when the SPM tip is placed over these
regions, only the flow from particular channels of the QPC are changed,
while other channels are not affected. This phenomenon is clearly
manifested in Fig.3, where the quantized $G\sim V_{g}$ curves of the QPC
with or without the SPM tip are plotted. From the comparison between the
behaviors of the solid line and the dash line in Fig.3(a), we can see that
when the tip is placed at the central axis of the QPC where the electrons
from the first mode travel through, the heights of
all plateaus are reduced. But the height difference
between each pair of neighboring plateaus
remains to be $2e^{2}/h$, except for that between the first plateaus
and the $0$-axis. In Fig.3(b), the SPM tip is positioned off
the central axis. According to Fig.2(a) and (b), we can learn that this
position corresponds to $\Delta G\sim 0$ in the first mode, and it
only affects the second mode. Thus in Fig.3(b) it is shown that the first
conductance plateau is almost unchanged, and the second and other higher
plateaus are lowered. Again, the height difference
between the second and the third
plateaus still remains to be $2e^{2}/h$.

\begin{figure}
\epsfxsize=7.5cm
\epsfbox{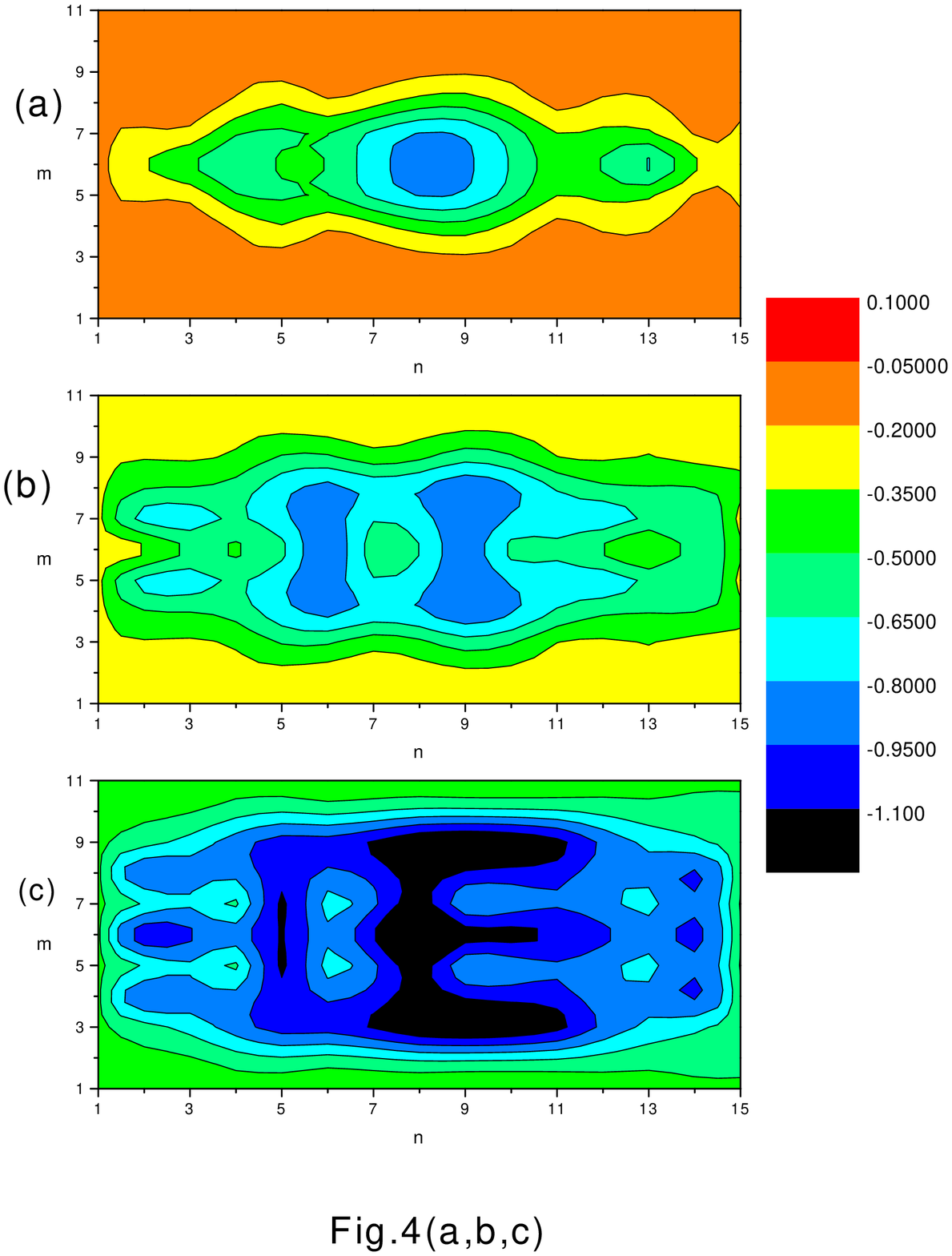}
\epsfxsize=7.5cm
\epsfbox{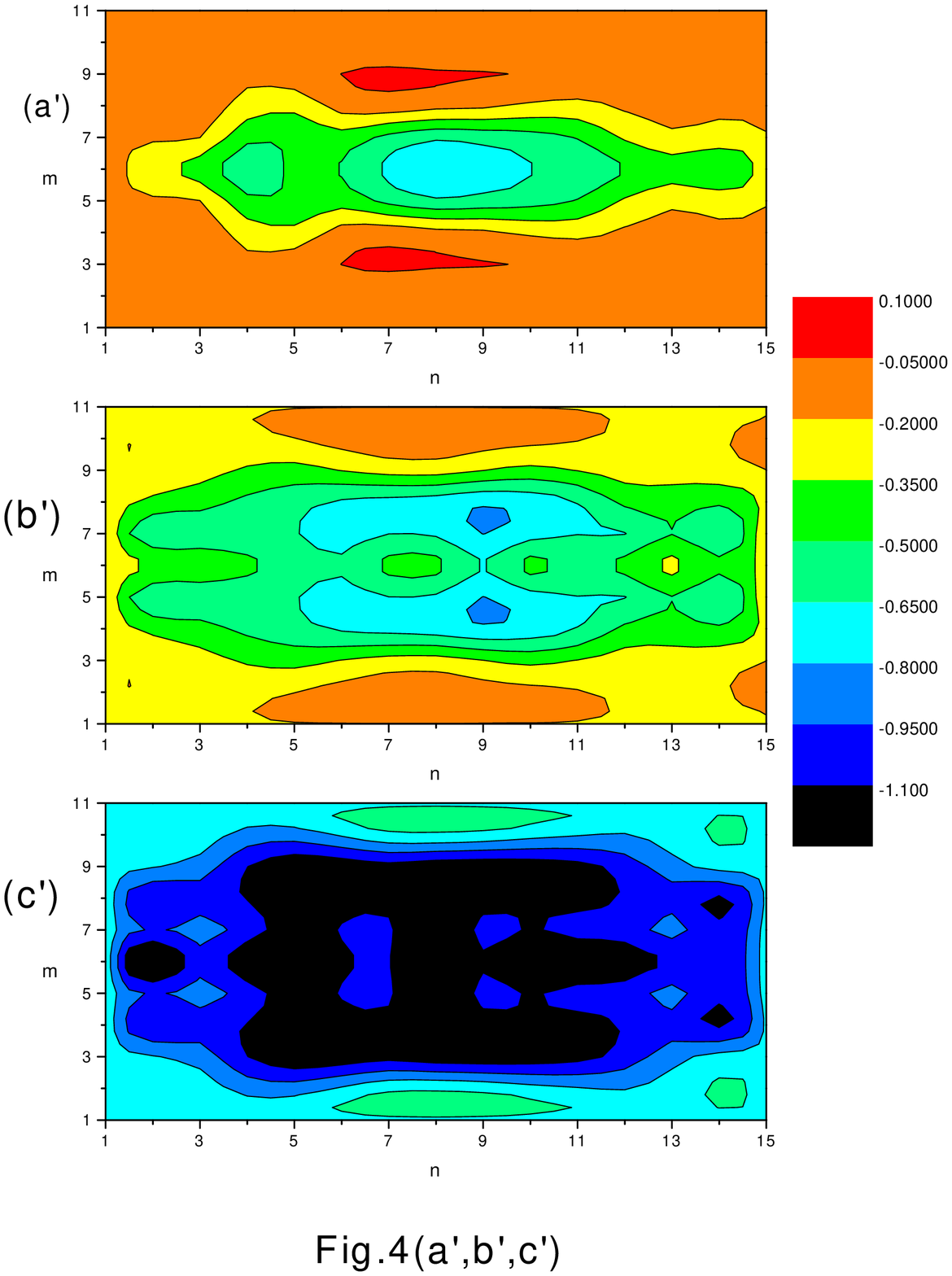}
\caption{Effect of electron heating on images of flow.
(a-c) show electron flow for the first, second
and third conductance plateaus under the voltage $eV=0.05t$.
$(a'-c')$ show corresponding images for higher voltage
$eV=0.2t$. Other parameters are the same as those in Fig.2.}
\end{figure}

The electron flows under the finite voltages across the QPC
are imaged in Fig.4,
where $(a-c)$ are at a low voltage
$(eV=0.05t)$ and $(a'-c')$
are at a high voltage $(eV=0.2t)$.
Comparing it with Fig.2, we see that the pattern still remains the same
in each mode,
but the values of $\Delta G$
is no longer so severely fluctuated when the voltage is raised.
A more important characteristic of voltage effect
is that fringe structure
would gradually disappear as the voltage rises.
Actually, we find that the three conductance modes
gradually merge into one conductance mode,
which implies that the quantized plateaus
also disappear as the voltage is increased.
The origin of the fringes is the coherent quantum
interference.\cite{Topinka_S}
Since the fringes
are less perceptible as the fluctuation of $\Delta G$ becomes pacified,
it is manifested that the consequence of using effectively hotter electrons
is to smear out the coherence among electrons. 
The above properties
are similar for elevating temperature
as the physical effects of
temperature and voltage
are the same here.\cite{Wees_91}
Clearly, the voltage effect addressed here
is in qualitative agreement with
the experimental observation.\cite{Topinka_S}

In conclusion, we have calculated the conductance of a QPC
in the presence of
the SPM tip by the recursive Green's function technique.
The numerical results are able to explain the
experimental results in Refs.7,8.

This work was supported in part by the RGC grant of Hong Kong
under Grant No. HKU7118/00P. 
G. P. H. was supported in part by the NSF of Guangdong Province 
(Contract No.011151) and the Foundation of Zhongshan University 
Advanced Research Centre (Contract No.02P2). 
Z. D. W. also thanks support in part from
the Texas Center for Superconductivity at the University of Houston.
S. L. Z. was supported in part by the
NSF of Guangdong under
Grant No. 021088.

\end{document}